\documentclass[aps,pra,reprint,showpacs]{revtex4-1}
\usepackage{graphicx}

\begin{document}     

\title{Efficient channeling of fluorescence photons from single quantum dots into guided modes of optical nanofiber}

\author{Ramachandrarao Yalla,$^{1}$ Fam Le Kien,$^{1}$ M. Morinaga,$^{2}$ and K. Hakuta$^{1}$}

\affiliation{$^{1}$Center for Photonic Innovations, University of Electro-Communications, Chofu, Tokyo 182-8585, Japan \\
$^{2}$Institute for Laser Science, University of Electro-Communications, Chofu, Tokyo 182-8585, Japan}

\date{\today}

\begin{abstract}
We experimentally demonstrate the efficient channeling of fluorescence photons from single q-dots on optical nanofiber into the guided modes, by measuring the photon-count rates through the guided and radiation modes simultaneously. We obtain the maximum channeling efficiency to be $22.0$ ($\pm4.8$)\% at fiber diameter of $350$ nm for the emission wavelength of $780$ nm. The results may open new possibilities in quantum information technologies for generating single photons into single-mode optical-fibers.  
\end{abstract}


\maketitle



Efficient collection of fluorescence photons from a single emitter into a single-mode fiber is a major challenge in the context of quantum information science. For that purpose various novel techniques have been proposed so far. The examples would include micropillar cavities \cite{yamamoto}, photonic crystal cavities \cite{vuckovic}, solid immersion lens \cite{sil}, and plasmonic metal nano-wires \cite{lukin}. However, in these techniques the subsequent coupling of fluorescence photons into a single-mode fiber may reduce the actual collection efficiency. In the view of the ability to directly couple fluorescence photons into a single-mode fiber, tapered optical fibers with sub-micron diameter, termed as optical nanofibers,  would be particularly promising. It has been theoretically predicted that, by positioning the emitter on the nanofiber surface, one can channel the fluorescence photons into the nanofiber guided-modes with an efficiency higher than $20\%$ \cite{ducloy, spem}, and moreover, fibers can be tapered adiabatically to keep the light transmission into the single-mode fiber higher than $90\%$ \cite{Nayak1, Arno1}.   

In the last decade, optical nanofibers  have been attracting considerable attention in the field of quantum optics. Many works have been reported so far using laser-cooled atoms. Channeling of fluorescence photons  into the guided modes has been demonstrated \cite{Nayak1}, and photon correlations from single atoms have been measured systematically through nanofiber guided modes \cite{Nayak2, Nayak3}. Fluorescence emission spectrum has been measured for few atoms through the guided mode by combining optical-heterodyne and photon-correlation methods \cite{Das}. Various schemes have been proposed for trapping atoms around the nanofiber \cite{trap0, trap, trap-Arno}, and the trapping has been experimentally demonstrated \cite{Arno2} using dipole-trapping method via two-color laser-fields \cite{trap}. However, regarding the channeling efficiency of fluorescence photons into the guided modes, it has not been measured yet, although the works so far imply a reasonable correspondence to the theoretical predictions \cite{Nayak1}. One reason would be due to a fact that atoms are not on the nanofiber surface and the atom-surface distance could not be estimated accurately.

Recently, two groups have reported the photon-counting measurements from semiconductor q-dots deposited on nanofibers \cite{nano, opex}. They absolutely measured the photon-count rates into the guided modes for one q-dot. They discussed the channeling efficiency of fluorescence photons into the guided modes based on the measured results. However, as pointed out in Ref. \cite{opex}, the value which can be obtained from such measurements is not the channeling efficiency itself, but  is a product of the channeling efficiency and the quantum efficiency of the q-dot. Therefore, the channeling efficiency cannot be determined from the measurements without accurate information on the quantum efficiency for the one q-dot which is measured. 
 
In the present Letter, we experimentally determine the channeling efficiency of fluorescence photons from single q-dots on optical nanofiber into the guided modes. We measure the photon-count rates through the guided and radiation modes simultaneously for various diameters of nanofiber. The measured results completely reproduce the theoretical predictions  \cite{ducloy, spem} within the experimental errors. The maximum channeling efficiency is obtained to be $22.0$ ($\pm4.8$)\% at the fiber diameter of $350$ nm for the emission wavelength of $780$ nm.

Figure 1 illustrates the schematic diagram of the experimental setup. Main part of the setup consists of inverted microscope (Nikon, Eclipse Ti-U) with a computer controlled $\it{x}$-$\it{y}$ stage, optical nanofiber, and sub-pico-liter needle-dispenser (Applied Micro Systems, ND-2000). Optical nanofiber is placed on the $\it{x}$-$\it{y}$ stage to precisely control the nanofiber position to the focus point of the microscope. Optical nanofibers are produced by adiabatically tapering commercial single-mode optical-fibers (SMF1, cut-off wavelength: $1.3$ $\mu$m) using a heat and pull technique. The diameter of nanofiber is measured using a scanning electron microscope (SEM) prior and after the optical experiments. The thinnest diameter of the nanofiber is $300$-$400$ nm, and  the nanofiber diameter varies along the fiber axis by $100$ nm/$1$ mm. The transmission through the optical nanofiber is measured to be $90\%$ using a fiber-coupled superluminescence light emitting diode (SLED) at $800$ nm.

We use core-shell type colloidal CdSeTe (ZnS) q-dots having emission wavelength at $790$ nm (Invitrogen, Q21371MP). We use the sub-pico-liter needle-dispenser  to deposit q-dots on nanofibers. The dispenser consists of a taper glass-tube which contains diluted q-dot solution and a needle having a tip of diameter $17$ $\mu $m. The needle axis is adjusted to coincide with the axis of the microscope, and the needle-tip position is computer-controlled along the axis. Once the needle tip passes through the taper glass-tube, it carries a small amount of q-dot solution at its edge. In order to deposit q-dots on nanofiber with minimum scattering loss, the needle-tip position is adjusted so that the q-dot solution at its tip just touches the nanofiber surface. Note that this method could deposit q-dots only on the $\it{upper}$ $\it{surface}$ $\it{of}$ $\it{nanofiber}$. The deposition is done for several positions along the fiber axis corresponding to the fiber diameter of $300$-$800$ nm. The transmission through the optical nanofibers is dropped to $81\%$ after the depositions. The depositions are done for three optical nanofiber samples, and the following measurements are carried out for all the deposited positions.

\begin{figure}
\centering\includegraphics[width=8cm]{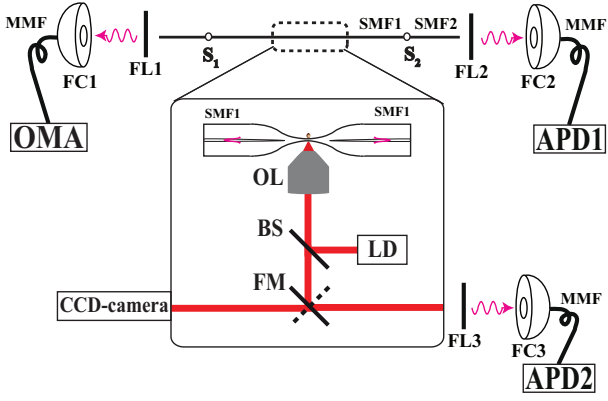}
\caption{(Color online) Schematic diagram of the experimental setup. Inset shows the optical nanofiber and microscope system. OL, BS, FM, and FL denote objective lens, beam splitter, flipper mirror, and filter, respectively. SMF, FC, and MMF denote single-mode fiber, fiber-coupler, and multi-mode fiber, respectively. LD, APD, and OMA denote laser diode, avalanche photodiode, and optical multichannel analyzer, respectively.}
\end{figure}

The q-dots are excited using cw laser-diode LD at a wavelength of $640$ nm. The excitation beam is focused to the nanofiber by the microscope objective lens OL ($40$X, NA= $0.6$). Regarding the fluorescence photons channeled into the guided modes, in order to guarantee the observation through the fundamental-mode ($HE_{11}$), SMF1 is spliced to another single-mode fiber SMF2 (cut-off wavelength: $557$ nm) at both ends marked as $S_{1}$, $S_{2}$ in Fig.1. The fluorescence light beam from each end of SMF2 is filtered from the scattered excitation laser light with a color glass filter FL1 (FL2) (HOYA, R72) and re-coupled into a multi-mode fiber. At one end of the  multi-mode fiber, fluorescence photons are detected with a fiber-coupled avalanche-photodiode APD1 (Perkin Elmer, SPCM-AQR/FC). At the other end of multi-mode fiber, fluorescence emission spectrum is measured using an optical-multichannel-analyzer OMA (Andor, DV420A-OE). 

Regarding the radiation modes, fluorescence photons are collected by OL, coupled into a multi-mode fiber by FC3, and detected by a fiber-coupled avalanche-photodiode APD2. A set of two filters FL3 (HOYA, R70/R72) is used to reject the scattered laser light from the focus point. Characteristics of APD1 and APD2 are the same, and signals from APD1 and APD2 are accumulated and recorded using photon-counting system (Hamamatsu, M8784). Photon-counting measurements for both guided and radiation modes and spectrum measurements are carried out for each deposited position simultaneously. Additionally, we perform photon-correlation measurements through the guided modes for all deposited positions \cite{opex}. All the above fluorescence measurements are carried out for the three nanofiber samples by keeping the excitation laser intensity at a low value of  $50$ W/cm$^{2}$ so that q-dots may not deteriorate \cite{degrade, blink}. 

The channeling efficiency $\eta_{c}$ into the nanofiber guided modes can be expressed as following,
\begin{equation}
\eta_{c}= \frac{n_{g}}{n_{g}+n_{r}}= \frac{1}{1+n_{r}/n_{g}}
\end{equation}
where $n_{g}$ and $n_{r}$ are photon emission rates into the guided and radiation modes, respectively. Observable photon-count rates by APD1 and APD2 are expressed as followings,
\begin{equation}
n_{g}^{(obs)}=\frac{1}{2}\eta_{APD1}\kappa_{g}n_{g},\ 
n_{r}^{(obs)}= \eta_{APD2}\kappa_{r}\eta _{r}n_{r}
\end{equation} 
where $\kappa_{g}$ and $\kappa_{r}$ are light-transmission factors for the paths of guided and radiation modes, respectively. Factor $1/2$  for $n_{g}^{(obs)}$ corresponds to a fact that fluorescence photons into the guided modes are detected only for one direction of the nanofiber. $\eta_{APD1}$ and $\eta_{APD2}$ are quantum efficiency of APD1 and APD2, respectively, and are assumed to be the same. $\eta_{r}$ is an effective collection efficiency for the radiation modes. Thus, the ratio ${n_{r}}/{n_{g}}$ can be written as following,
\begin{equation}
\frac{n_{r}}{n_{g}}= \frac{n_{r}^{(obs)}}{n_{g}^{(obs)}}\times{\frac{\kappa_{g}}{2\kappa_{r}\eta _{r}}}=  \frac{n_{r}^{(obs)}}{n_{g}^{(obs)}}\times C
\end{equation}
where $C$= ${\kappa_{g}}/{{2\kappa_{r}\eta _{r}}}$.

$\kappa_{g}$-value was measured to be $49.6$ ($\pm2.1$)\%. The measurement procedure is following. The SLED output is spliced to SMF2 at the FL1-end, and the output power is measured at the APD1-position.  Input power to the optical nanofiber is measured by cleaving the SMF1 before entering into the optical nanofiber. The measured value is consistent with a value calculated as a product of transmission factors of  optical nanofiber ($81\%$), splicing point between SMF1 and SMF2 ($81\%$), FL2 ($83\%$), and coupling efficiency into the multi-mode fiber at FC2 ($90\%$). 

$\kappa_{r}$-value was obtained to be $23.5$ ($\pm1.3$)\% as a product of transmission factors of all optical components in the path and coupling efficiency into multi-mode fiber. Transmission factors are measured for OL ($74\%$), BS ($63\%$), FM ($83\%$), and FL3 ($75\%$), using the SLED light. The coupling efficiency into multi-mode fiber at FC3 was obtained to be $81\%$ by the following procedure. First,  SLED light is  introduced from the LD-port and is focused to the nanofiber. The scattered light from the focused spot is collected through the OL, and its power is measured both at FC3-position and at APD2-position through multi-mode fiber.

Regarding the radiation modes, the effective collection efficiency $\eta_{r}$ consists of two factors. One is from numerical aperture (NA) of the OL. The collection efficiency of the OL is estimated to be $10\%$ from the NA-value of $0.6$. The other factor arises from the nanofiber itself. The q-dots are deposited on the upper surface of nanofiber and the OL collects the fluorescence photons from the down side of nanofiber. Therefore, the nanofiber acts as a cylindrical lens and the collection efficiency of the OL may be enhanced by the lens effect of nanofiber. We calculated the enhancement factor based on the formalism developed in Ref. \cite{Stratton}, and estimated the average enhancement factor by assuming random azimuthal distribution of q-dots on the upper surface of nanofiber. It was found that the average enhancement factor could be assumed to be constant with a value of $1.48$ ($\pm0.03$) for the fiber diameters of the present measurements. We use this average enhancement factor to obtain the effective collection efficiency $\eta_{r}$. Thus, we obtain the $\eta_{r}$-value to be $14.8$ ($\pm0.3$)\% and consequently the $C$-value to be $7.13$ ($\pm0.84$) by combining the values of $\kappa_{g}$, $\kappa_{r}$, and $\eta_{r}$.

\begin{figure}
\centering\includegraphics[width=8cm]{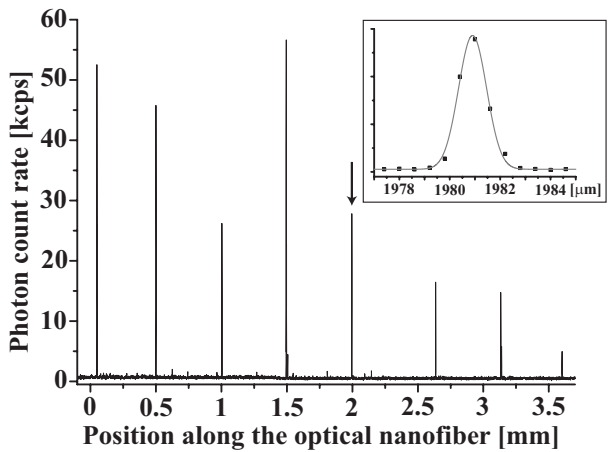}
\caption{(Color Online) Observed fluorescence photon-count rate by scanning the focusing point along the nanofiber. Origin of the horizontal axis corresponds to the center of nanofiber. Inset shows an expanded profile of a peak marked by arrow. Gray curve shows a Gaussian fitting with a width of $1.5$ $\mu $m FWHM. }
\end{figure}
Figure 2 shows the fluorescence photon-count rate measured for a deposited nanofiber by scanning the focusing point along the nanofiber. The scanning speed is $6$ $\mu $m/s, and signals are measured through the guided mode by APD1. One can clearly see eight sharp peaks along the nanofiber with a typical separation of $0.5$ mm. Origin of the horizontal axis corresponds to the center of the nanofiber. Nanofiber diameter varies from 400 nm at the origin to 750 nm at the position $3.5$ mm. Inset shows an expanded profile of a peak marked by arrow. The width should be limited by the focused spot size on nanofiber and is about $1.5$ $\mu $m FWHM.

\begin{figure}
\centering\includegraphics[width=8cm]{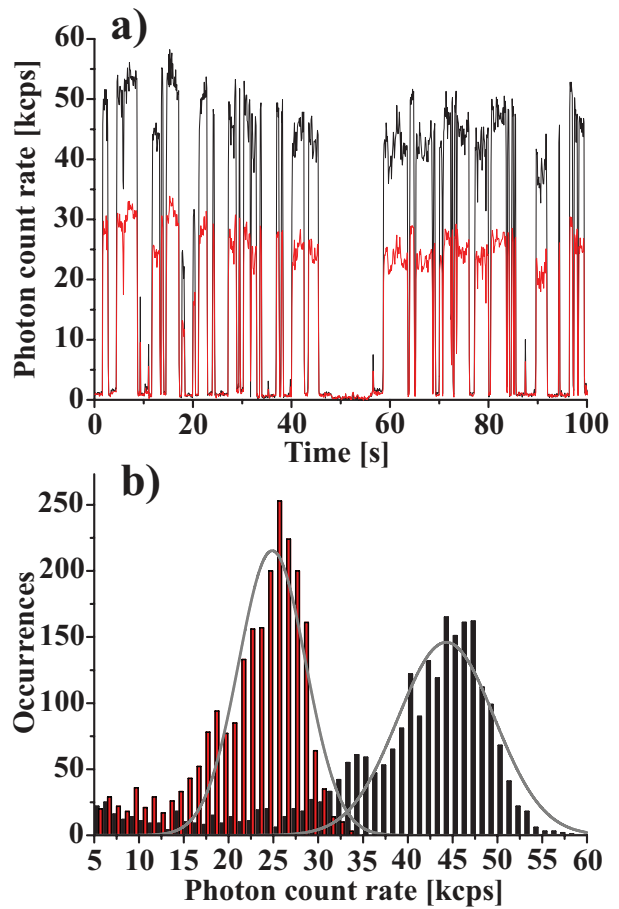}
\caption{(Color Online) (a) Typical fluorescence photon-count rate from a single q-dot as a function of time. Black and red traces correspond to fluorescence photon-count rates observed through guided and radiation modes, respectively. (b) Histograms for the count rates plotted for guided and radiation modes. Black (red) histograms correspond to the guided (radiation) modes. Fittings by Gaussian profiles are drawn by gray solid curves.}
\end{figure}

 Figure 3(a) shows the typical fluorescence photon-count rates from q-dots on nanofiber at fiber diameter of 400 nm. Black and red traces correspond to the photon-count rates through the guided and radiation modes, respectively. Measurement time is $5$ min with a time bin size of $100$ ms. One can readily see that the two traces exactly match by each other. Photon-count rates show a clear single step blinking behavior, revealing that the number of deposited q-dots is one. This single q-dot deposition could be further confirmed by measuring the anti-bunching dip in the normalized photon-correlations, and the dip-value was measured to be $0.035$ $\textless\textless1$. Above measurements were performed for all the depositions of the three nanofiber samples, and the number of q-dots at each deposition was measured to be one or two, similarly as in Ref.  \cite{opex}. Regarding the fluorescence spectrum, the center wavelength distributes over the range of $80$ nm from $740$ to $820$ nm with a typical FWHM of 52 nm. 
 
Figure 3(b) shows the photon-count rate histograms for the black and red traces for the whole measurement time with a counting interval of $1$ kcps. By fitting the histograms with Gaussian profiles \cite{blink}, we obtain ${n_{g}^{(obs)}}$ and ${n_{r}^{(obs)}}$ to be  $44.3$ ($\pm5.4$) and $24.8$ ($\pm3.7$) kcps, respectively. Using the relation of Eq. (3), the ratio ${n_{r}}/{n_{g}}$ is  obtained to be  $3.99$ ($\pm1.55$). Thus, we obtain using the Eq. (1) the $\eta_{c}$-value to be $20.0$ ($\pm6.2$)\%. Using the same procedure, the $\eta_{c}$-values were obtained at various fiber diameters for the three nanofiber samples. We obtained average value of  $\eta_{c}$ at each fiber diameter for the three nanofiber samples. Regarding the fiber diameter of $400$ nm, the average $\eta_{c}$-value was $21.5$ ($\pm2.4$)\%.

\begin{figure}
\centering\includegraphics[width=8cm]{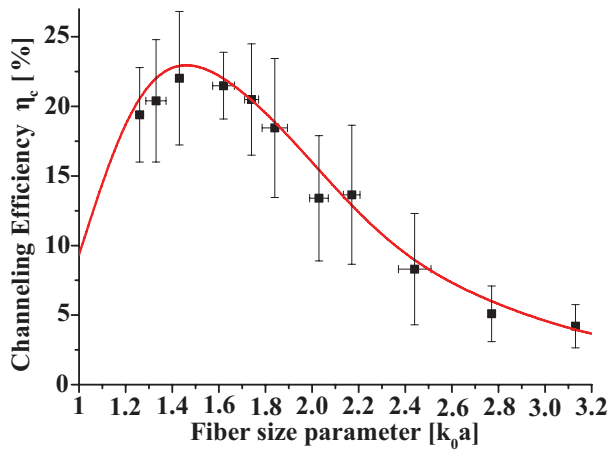}
\caption{(Color Online) Channeling efficiency as a function of fiber size parameter ($k_{0}$a= $2$$\pi$$a$/$\lambda$). Red curve denotes the theoretical prediction. Measured values are marked by black squares with error bars. }
\end{figure} 

Figure 4 shows the channeling efficiency $\eta_{c}$ as a function of fiber size parameter ($k_{0}$a= $2$$\pi$$a$/$\lambda$). The size parameter is calculated for each deposited position by using the measured fiber-diameter $2a$ and the observed emission-wavelength $\lambda$. The red curve exhibits the theoretical prediction for the channeling efficiency into the $HE_{11}$-mode assuming the nanofiber refractive-index of 1.45. Experimental values are shown by black squares with error bars. The error bars in horizontal axis are due to the variation of emission wavelength at each deposited position. The error bars in vertical axis are due to the fluctuation of the measured values. The fluctuation would mainly be due to the measurement ambiguity, but another cause may happen from the ambiguity of the enhancement factor due to the nanofiber lens effect. For the experimental analysis, we used the average enhancement factor assuming random azimuthal distribution of deposition, but the enhancement factor for each deposited position would be different from the average value. Such ambiguity should induce the fluctuation in the measured results. Although ambiguity of $\pm20$\% still exists, attention must be specially paid that the measured results have completely reproduced the theoretical prediction. The channeling efficiency into the guided modes reaches to a maximum value of  $22.0$ ($\pm4.8$)\% at the fiber size parameter of  $1.43$, which corresponds to the fiber diameter of $350$ nm for the emission wavelength of $780$ nm. 


In summary, we have experimentally demonstrated the efficient channeling of fluorescence photons from single q-dots on optical nanofiber into the guided modes, by measuring the photon-count rates through the guided and radiation modes simultaneously. We have obtained the maximum channeling efficiency to be $22.0$ ($\pm4.8$)\% at the fiber diameter of $350$ nm for the emission wavelength of $780$ nm. The present results may open new possibilities in quantum information technologies for generating single photons into single-mode optical-fibers.


We thank Kali Nayak for helpful discussions. This work was carried out as a part of the Strategic Innovation Project by Japan Science and Technology Agency.

\end{document}